\pdfoutput=1

\documentclass{article}
\usepackage{natbib} 
\usepackage[utf8]{inputenc}
\usepackage{gensymb}
\usepackage{hyperref}
\usepackage{graphicx} 
\usepackage{subcaption} 
\usepackage{amsmath}
\usepackage{units} 
\usepackage[a4paper]{geometry}
\usepackage[labelfont=bf]{caption}

\usepackage[flushmargin]{footmisc}

\title{Effects of Scan Length and Shrinkage on Reliability of Resting-State Functional Connectivity in the Human Connectome Project}
\author{Amanda F. Mejia, Mary Beth Nebel, Anita D. Barber, Ann S. Choe and Martin A. Lindquist}
\date{}

\begin{document}

\maketitle

\begin{abstract}
In this paper, we use data from the Human Connectome Project ($\text{N}=461$) to investigate the effect of scan length on reliability of resting-state functional connectivity (rsFC) estimates produced from resting-state functional magnetic resonance imaging (rsfMRI).  Additionally, we study the benefits of empirical Bayes shrinkage, in which subject-level estimates borrow strength from the population average by trading a small increase in bias for a greater reduction in variance. For each subject, we compute raw and shrinkage estimates of rsFC between 300 regions identified through independent components analysis (ICA) based on rsfMRI scans varying from $3$ to $30$ minutes in length. The time course for each region is determined using dual regression, and rsFC is estimated as the Pearson correlation between each pair of time courses. Shrinkage estimates for each subject are computed as a weighted average between the raw subject-level estimate and the population average estimate, where the weight is determined for each connection by the relationship of within-subject variance to between-subject variance. We find that shrinkage estimates exhibit greater reliability than raw estimates for most connections, with 30-40\% improvement using scans less than 10 minutes in length and 10-20\% improvement using scans of 20-30 minutes. We also observe significant spatial variability in reliability of both raw and shrinkage estimates, with connections within the default mode and motor networks exhibiting the greatest reliability and between-network connections exhibiting the poorest reliability.  We conclude that the scan length required for reliable estimation of rsFC depends on the specific connections of interest, and shrinkage can be used to increase reliability of rsFC, even when produced from long, high-quality rsfMRI scans.
\end{abstract}

\section{Introduction}

A growing concern in psychological science is measurement reliability \citep{open2015estimating, button2013power,munafo2014scientific}, including the reliability of inter-individual differences in functional connectivity within the brain as measured using resting-state functional magnetic resonance imaging (rsfMRI) \citep{shehzad2009resting}. One practice that has been shown to improve reliability of subject-level rsFC is empirical Bayes shrinkage, in which subject-level observations ``borrow strength'' from a larger group of subjects. Empirical Bayes shrinkage estimates are a weighted combination of a subject-level observation and the average over a group of subjects, where the degree of shrinkage towards the group average depends on the reliability of the subject-level observation and the similarity of subjects in the group.  In \cite{mejia2015improving}, we considered voxel-level rsFC produced using relatively short (5-7 minute) scans and showed that performing shrinkage resulted in improvement in reliability of rsFC by $25$-$30\%$.  While this provided clear evidence for the benefits of shrinkage for voxel-level rsFC in cases when longer scans are not feasible, it was not clear whether the benefits would carry over in different circumstances, such as increased temporal resolution and scan duration. 

The duration of rsfMRI scans plays an important role in determining the reliability of rsFC estimates, as sampling variability decreases with increased scan length.  Furthermore, as subject-level rsFC varies due to changes in the cognitive and emotional state of the subject \citep{shehzad2009resting, birn2013effect}, longer scans sample over a wider variety of cognitive states of the subject, providing a better estimate of the subject's average rsFC.  Similarly, combining estimates of rsFC from multiple sessions may produce more reliable estimates of rsFC than using data from a single scan of the same (combined) duration \citep{shehzad2009resting, laumann2015functional}. Several studies have focused on determining the scanning duration needed to reliably estimate rsFC \citep{van2010intrinsic, anderson2011reproducibility, birn2013effect} and related measures \citep{van2010intrinsic, kalcher2012fully, hacker2013resting, zuo2013toward, li2014influence, whitlow2011effect, liao2013functional, murphy2007long}.  While increased scan duration has been consistently found to improve reliability, different studies have reached vastly different conclusions about what scan duration is sufficient, with recommendations ranging from 5 minutes \citep{whitlow2011effect, liao2013functional}, to 90 minutes or more \citep{laumann2015functional}.  This discrepancy may be attributed to (i) differences in the definition and metric of reliability and (ii) the fact that there are many other factors besides scan length that influence reliability.

In terms of the definition of reliability, studies may examine intersession reliability, intrasession reliability, or end-point reliability.  \textit{Intersession reliability} asks, how similar is the rsFC of a subject across multiple scanning sessions occurring days or even months apart \citep{choe2015reproducibility} and is usually the true measure of interest.  \textit{Intrasession reliability} asks, how similar is the rsFC of a subject across multiple runs within the same session.  This tends to overestimate true intersession reliability \citep{shehzad2009resting, anderson2011reproducibility, birn2013effect, zuo2013toward}, but can serve as a reasonable alternative when multiple rsfMRI sessions are not available.  In some cases, intrasession reliability may be of interest, for example for the study of state-level, rather than trait-level, effects \citep{geerligs2015state}.\footnote{State-level characteristics are temporary and vary within an individual, such as tiredness or surprise, while trait-level characteristics are relatively stable over time and vary across individuals, such as IQ or temperment \citep{robins2009handbook}.}  
\textit{End-point reliability} asks, how similar is an estimate of rsFC produced from a portion of a scanning session to the estimate produced using the full session.  This is not a true measure of reliability in the strictest sense, as the estimates being compared are not independent, and it tends to overestimate true intersession or intrasession reliability.  
Finally, studies differ in terms of the metric used to quantify reliability, which include intra-class correlation coefficient (ICC) \citep{shehzad2009resting, birn2013effect, zuo2013toward, guo2012one, chou2012investigation}, correlation \citep{shehzad2009resting, van2010intrinsic, laumann2015functional}, and mean squared or absolute error \citep{anderson2011reproducibility, birn2013effect}. 

Factors other than scan length that may influence reliability of rsFC include factors related to acquisition, preprocessing and analysis.  In terms of acquisition, increasing temporal resolution \citep{birn2013effect, zuo2013toward, liao2013functional} and having subjects lie with eyes open rather than closed \citep{van2010intrinsic} have been shown to result in improved reliability.  
In terms of preprocessing, several studies have found that global signal regression tends to worsen reliability \citep{zuo2013toward, liao2013functional}, while nuisance regression tends to improve reliability \citep{zuo2013toward}.  In addition, performing analysis in surface rather than volumetric space may result in improved reliability \citep{zuo2013toward}.  In terms of rsFC analysis, the use of functional versus anatomical regions of interest (ROIs) \citep{anderson2011reproducibility} and accurate identification of functional ROIs \citep{smith2011network} have been found to improve reliability.  
Reliability has also been shown to vary with ROI size \citep{hacker2013resting} and the specific connections being studied \citep{shehzad2009resting, van2010intrinsic, anderson2011reproducibility, laumann2015functional, mueller2015reliability}.  For example, higher reliability has been observed for correlations that are statistically significant at the group level; for within-network versus between-network correlations; and for connections within the DMN network versus within the task positive, attention, motor and visual networks \citep{shehzad2009resting, van2010intrinsic, laumann2015functional}.

The question of how long to scan in order to produce sufficiently reliable rsFC estimates is clearly a difficult one and depends on many factors, including the specific connections of interest.  It therefore remains important to both increase the duration and/or number of scans whenever possible and to adopt best practices for improving reliability, including shrinkage.  In this paper, we use the Human Connectome Project (HCP) to examine the effects of scan length and shrinkage on reliability of whole-brain rsFC between functional ROIs identified through ICA.  The rsfMRI scans of the HCP are relatively long (~60 minutes over two visits), are produced using an advanced acquisition and preprocessing pipeline, have been ``denoised'' to remove sources of non-neuronal variation, and have been transformed to surface space.  Therefore, the HCP provides an opportunity to determine the effect of shrinkage on rsFC reliability when many other best practices, including increased scan length, have already been adopted.


\section{Methods}

\subsection{Data and connectivity estimation}

\subsubsection{Human Connectome Project Data}
\label{sec:data}
The Human Connectome Project (HCP) \citep{van2013wu} is a collection of neuroimaging and phenotypic information for over one thousand healthy adult subjects, which are being incrementally released to the scientific community (\url{http://humanconnectome.org}). For the analyses described below, we used the following data from the 523 subjects included in the 2014 Human Connectome Project 500 Parcellation+Timeseries+Netmats (HCP500-PTN) release. All MRI data were acquired on a customized 3T Siemens connectome-Skyra 3T scanner, designed to achieve 100 mT/m gradient strength. For 461 of the 523 subjects, a multi-band / multi-slice pulse sequence with an acceleration factor of eight \citep{moeller2010multiband, feinberg2010multiplexed, setsompop2012blipped, xu2012highly, uugurbil2013pushing} was used to acquire four roughly 15-minute rsfMRI sessions, each consisting of 1200 volumes sampled every 0.72 seconds at 2 mm isotropic spatial resolution. The sessions were collected over two visits that occurred on separate days, with two runs collected at each visit. Across sessions at each visit, phase encoding directions were alternated between right-to-left (RL) and left-to-right (LR) directions.  Before October 1, 2012, the first run of each visit was acquired with RL phase encoding, and the second run was acquired with LR phase encoding (RL/LR).  After this date, the first visit continued to be acquired in the RL/LR order, but the second visit was acquired in the opposite order, with the LR acquisition followed by the RL acquisition (LR/RL).

Spatial preprocessing was performed using the minimal preprocessing pipeline as described by \cite{glasser2013minimal}, which includes correcting for spatial distortions and artifacts, and projection of the data time series to the standard grayordinate space. Structured artifacts in the time series were removed using ICA + FIX (independent component analysis followed by FMRIB's ICA-based X-noiseifier; \citeauthor{salimi2014automatic}, \citeyear{salimi2014automatic}; \citeauthor{griffanti2014ica}, \citeyear{griffanti2014ica}), and each data set was temporally demeaned with variance normalization according to \cite{beckmann2004probabilistic}.  Group independent component analysis (GICA) was performed on the full rsfMRI time series for all 461 subjects to estimate a set of spatial independent components (ICs) that represent population-average resting-state networks \citep{beckmann2004probabilistic}.  GICA was performed using model orders of 25, 50, 100, 200 and 300 independent components (ICs).  After identification of spatial ICs at each model order, time courses were estimated for each subject and IC by performing the first stage of dual regression \citep{beckmann2009group}.  Specifically, the group IC spatial maps were used as predictors in a multivariate linear regression model against the full rsfMRI time series, which was created by concatenating the four sessions of each subject into a single time series (of length $4800$ volumes) in the following order: visit 1 LR, visit 1 RL, visit 2 LR, visit 2 RL (see Figure \ref{fig:data}). 

\subsubsection{Connectivity matrix estimation}
\label{sec:conn_estimation}

The quantity of interest for each subject is the true $Q\times Q$ connectivity matrix, representing the pairwise connectivity during rest between each of the $Q\in\{25, 50, 100, 200, 300\}$ regions identified through GICA.  We use Pearson correlation as our measure of connectivity.  We are interested in how the reliability of estimates of this connectivity matrix changes with longer scan duration.  To this end, we estimate the connectivity matrix using the first $\ell$ volumes of the time series from visit 1 for each subject, with $\ell\in\{300,600,\dots,2400\}$.  With a TR of $0.72$ seconds, the resulting time series range from $3.6$ to $28.8$ minutes in duration.  We also estimate the connectivity matrix using all $L=2400$ volumes of visit 2 for each subject.


\textit{Shrinkage of connectivity estimates.}
Shrinkage estimators, which ``borrow strength'' from the population to improve subject-level estimates, have been shown to improve reliability of voxel-level connectivity estimates based on short rsfMRI scans \citep{shou2014shrinkage, mejia2015improving}.  Here, we assess the ability of shrinkage estimators to improve reliability of connectivity estimates produced from longer scans.  

We first provide a brief introduction to empirical Bayes shrinkage estimators, by focusing on a simple measurement error model \citep{carroll2006measurement}.  For subjects $i=1,\dots,n$, let the true connectivity between two regions $q$ and $q'$ be denoted $X_i(q,q')$.  For visit $j$ and scan length $\ell$, we have an estimate of $X_i(q,q')$, which we denote $W_{ij}^{(\ell)}(q,q')$.  The measurement error model assumes that the estimate $W_{ij}^{(\ell)}(q,q')$ can be decomposed into a signal $X_i(q,q')$ and a noise term $U_{ij}^{(\ell)}(q,q')$:

\begin{equation}\label{eqn:meas_err}
W_{ij}^{(\ell)}(q,q') = X_i(q,q') + U_{ij}^{(\ell)}(q,q'),
\end{equation}

where $X_i(q,q')\sim N\{\mu(q,q'),\sigma^2_x(q,q')\}$ and $U_{ij}^{(\ell)}(q,q')\sim N\{0,\sigma^{2(\ell)}_u(q,q')\}$.  We assume that $X_i(q,q')$ and $U_{ij}^{(\ell)}(q,q')$ are independent, the $X_i(q,q')$ are independent across subjects, and the $U_{ij}^{(\ell)}(q,q')$ are independent across subjects and visits.  Then the shrinkage estimator of $X_i(q,q')$ is equal to the empirical posterior mean,

$$
\tilde{W}_{ij}^{(\ell)}(q,q') = 
\lambda_t(q,q')\bar{W}_{\cdot j}^{(\ell)}(q,q') +
\left\{1-\lambda_t(q,q')\right\}W_{ij}^{(\ell)}(q,q'),
$$

where $\bar{W}_{\cdot j}^{(\ell)}(q,q')=\tfrac{1}{n}\sum_{i=1}^n W_{ij}^{(\ell)}(q,q')$. The shrinkage parameter $\lambda_t(q,q')$ is given by

$$
\lambda_t(q,q') = \frac{\sigma^{2(\ell)}_u(q,q')}{\sigma^{2(\ell)}_u(q,q')+\sigma^2_x(q,q')}
$$

and ranges from $0$ (no shrinkage) to $1$ (complete shrinkage to the group mean), depending on the relative size of the within-subject and between-subject variance terms.  Estimation of these variance components is, in theory, straightforward: defining $\sigma^{2(\ell)}_w(q,q')$ as the total variance, which is estimated as the variance of the $W_{ij}^{(\ell)}(q,q')$, averaged over visits, the noise variance can be estimated as 

$$
\hat\sigma^{2(\ell)}_u(q,q')=\frac{1}{2}Var_i\{ W_{i2}^{(\ell)}(q,q') - W_{i1}^{(\ell)}(q,q')\},
$$ 

and the signal variance can be estimated as 

$$
\hat\sigma^2_x(q,q') = \hat\sigma^{2(\ell)}_w(q,q') - \hat\sigma^{2(\ell)}_u(q,q').
$$

However, this presupposes the availability of multiple observations or visits for each subject.  While this is true in our case, it is not a reasonable assumption in general, as many studies only collect a single resting-state fMRI scan for each subject.  Furthermore, when multiple visits or sessions are available, they can be combined to create a single estimate of connectivity with improved accuracy \citep{laumann2015functional}.  Therefore, the salient problem is to estimate the within-subject variance of a \textit{single} connectivity estimate produced using \textit{all} of the rsfMRI data available for each subject.  This clearly precludes the availability of multiple observations of the quantity of interest.  In \cite{mejia2015improving}, we proposed a solution based on the idea of ``pseudo scan-rescan'' data, in which a single scanning session is treated as two sessions, composed respectively of the first and second halves of the time series.  This approach was also recently applied in the context of reliability correction by \cite{mueller2015reliability}.  In \cite{mejia2015improving}, we found the within-subject variance estimate produced from this approach to be upwardly biased, and proposed using an empirical adjustment factor to correct for this.  However, this adjustment method assumes that the majority of within-subject variance can be attributed to sampling variance (as this source of variance is inflated using pseudo scan-rescan data). It is therefore best suited to connectivity estimates produced from noisier time series as in the case of \cite{mejia2015improving}, where voxel-level connectivity based on short (7.5 minute) scans were estimated.  We have since developed a more general method of estimating within-subject variance from a single session, which seeks to separate sampling variance from other sources of within-subject variance.  While a full description of this method is beyond the scope of this paper, a brief description is provided in the Appendix.

We also assess the performance of an ``oracle'' shrinkage estimator, which uses both visits from each subject to estimate the variance components.  While this is not realistic (since, again, if multiple visits are available they would be combined into a single, improved estimator), it provides an upper bound on the performance of shrinkage estimators, since it is based on the best---if realistically unattainable---estimate of within-subject variance.  In the continuation we denote this estimate $\tilde{W}_{ij}^{*(\ell)}(q,q')$.

\subsection{Reliability of connectivity estimates}

We now describe the methods used to quantify reliability of the raw and shrinkage estimates of connectivity for each subject.  We are primarily interested in \textit{intersession reliability} but we also assess \textit{end-point reliability} to illustrate the bias inherent in this approach.

\begin{figure}
\centering
\includegraphics[page=1, scale=0.6, trim = 0in 4.2in 0.3in 1.2in, clip]{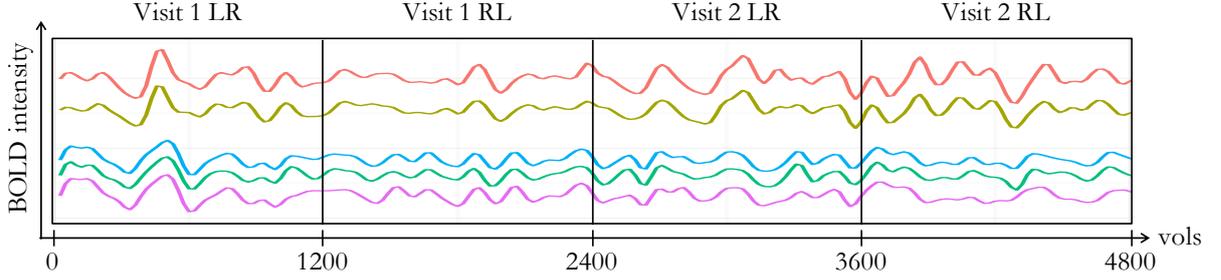}
\caption{Illustration of the data for a single subject.  For five regions, the full time series, consisting of $4800$ volumes, is shown.  Each time series consists of four sessions, occurring over two visits.  In the GICA provided in the HCP data release, the sessions were reordered so that both visits are concatenated in the LR/RL order.}
\label{fig:data}


\centering
\includegraphics[page=2, scale=0.6, trim = 0.5in 0in 1in 0in, clip]{Figures.pdf}
\caption{Illustration of data setup for intersession and end-point reliability analysis for a single subject $i$.  For intersession reliability analysis, we are interested in how similar the connectivity estimates $W_{i1}^{(\ell)}$, $\tilde{W}_{i1}^{(\ell)}$ and $\tilde{W}_{i1}^{*(\ell)}$ are to the full visit 2 raw estimate $W_{i2}^{(L)}$, $L=2400$, as $\ell$ varies from $300$ to $2400$.  For end-point reliability analysis, we are interested in how close the connectivity estimates $W_{i1}^{(\ell)}$, $\tilde{W}_{i1}^{(\ell)}$ and $\tilde{W}_{i1}^{*(\ell)}$ are to the full visit 1 raw estimate $W_{i1}^{(L)}$, as $\ell$ varies from $300$ to $2400$.}
\label{fig:setup}
\end{figure}

As illustrated in Figure \ref{fig:setup}a, in order to assess intersession reliability of the raw and shrinkage connectivity estimates for each subject, we compare the estimates produced using the first $\ell$ volumes of visit 1 to the raw connectivity estimate produced using all $L=2400$ volumes of visit 2.  As shown in Figure \ref{fig:setup}b, to assess end-point reliability for each subject, we compare the raw and shrinkage connectivity estimates produced using the first $\ell$ volumes of the first visit to the raw estimate produced using all $L$ volumes of the same visit.  

The metric we use to quantify reliability is absolute percent error (APE).  Specifically, the intersession reliability of the raw estimate of connectivity between regions $q$ and $q'$ for subject $i$ and scan length $\ell$ is 
$$
\text{APE}^{(\ell)}_{\text{raw},i}(q,q') = \left|\frac{W_{i1}^{(\ell)}(q,q') - W_{i2}^{(L)}(q,q')}{W_{i2}^{(L)}(q,q')}\right|,
$$ 
while the intersession reliability of the corresponding shrinkage estimate is 
$$
\text{APE}^{(\ell)}_{\text{shrink},i}(q,q') = \left|\frac{\tilde{W}_{i1}^{(\ell)}(q,q') - W_{i2}^{(L)}(q,q')}{W_{i2}^{(L)}(q,q')}\right|.
$$ 
End-point reliability of raw and shrinkage estimates is computed in a similar way by replacing $W_{i1}^{(L)}(q,q')$ by $W_{i2}^{(L)}(q,q')$.  

\begin{figure}
\centering
\includegraphics[page=4, scale=0.6, trim = 0.5in 0.3in 0.5in 0.3in, clip]{Figures.pdf}
\caption{We summarize reliability at three different resolutions: edge-level, seed-level and omnibus.  \textit{Edge-level reliability} is computed as the median reliability across subjects at each edge, resulting in a map of reliability for each seed region.  \textit{Seed-level reliability} is computed as the median edge-level reliability within each seed, resulting in a single map of reliability.  \textit{Omnibus reliability} is computed as the median edge-level reliability across all unique pairs of regions, resulting in a single scalar summary measure of reliability.\\[10pt]}
\label{fig:reliability_hierarchy}


\centering
\includegraphics[width=3.5in, trim=0 0 25cm 1cm, clip]{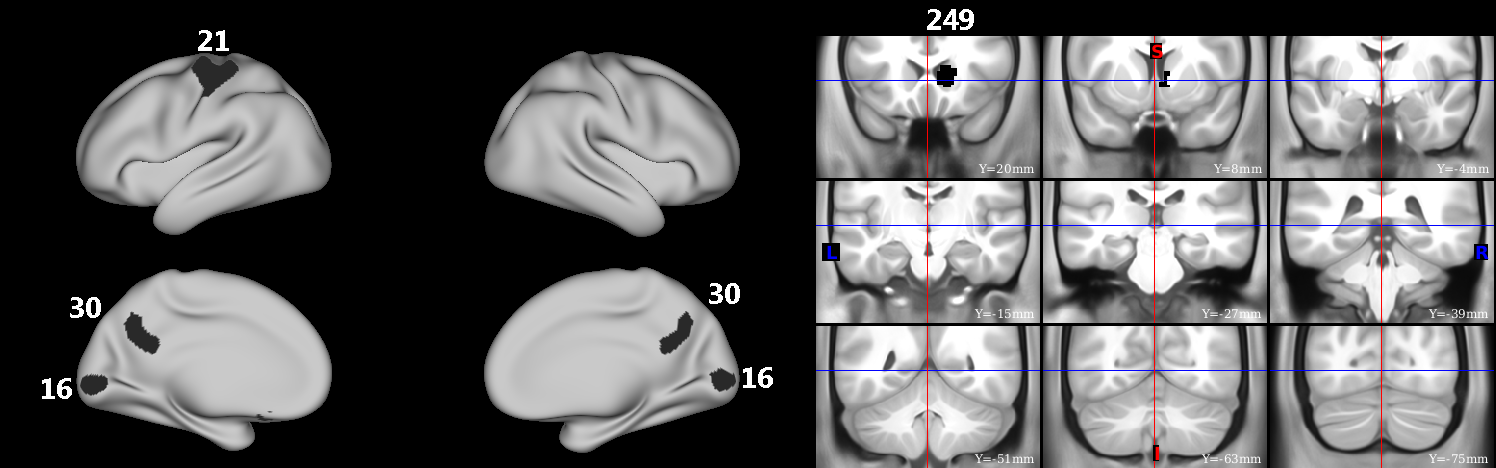}
\caption{Three selected seed regions selected from model order $300$, lying respectively in the visual cortex (IC 16), the somatomotor cortex (IC 21), and the DMN (IC 30).  The visual seed is located in the bilateral lingual gyrus; the somatomotor seed is located in the left dorsolateral pre- and post-central gyri; the DMN seed is located in the posterior cingulate cortex (PCC).}
\label{fig:seeds}
\end{figure}

As illustrated in Figure \ref{fig:reliability_hierarchy}, using APE as the measure of reliability, we summarize reliability over subjects at three different resolutions: edge-level, seed-level and omnibus.  This organization provides both a high-level view of how reliability changes with additional scan duration and the use of shrinkage estimates, and a detailed view of how reliability varies across different pairs of regions and how scan duration and shrinkage affect the reliability of specific connections.  We first compute the median reliability across all subjects for each edge (or pair of regions).  The result is \textit{edge-level reliability}, illustrated in the top panel of Figure \ref{fig:reliability_hierarchy}.  Edge-level reliability can be visualized as a set of images, each showing the reliability of connectivity between a single seed and all other regions in the brain.  As visualization of all regions and model orders is impractical, we select for visualization three seed regions lying within well-known resting state networks, including the visual cortex, the somatomotor cortex, and the default mode network (DMN) (Figure \ref{fig:seeds}).  Next, we compute \textit{seed-level} reliability by treating each region as a seed and computing the median edge-level reliability of connectivity with all other regions in the brain.  Illustrated in the middle panel of Figure \ref{fig:reliability_hierarchy}, seed-level reliability can be visualized as a single image, illustrating for each region the overall reliability of connectivity.  Finally, we compute \textit{omnibus} reliability as the median edge-level reliability across all unique pairs of regions, resulting in a single scalar summary of reliability as illustrated in the bottom panel of Figure \ref{fig:reliability_hierarchy}.



\section{Results}

This section is organized as follows: we begin by looking at omnibus reliability as a function of scan length and ICA model order to establish general trends, then ``zoom in'' to better understand more subtle patterns by examining seed-level and then edge-level reliability. At the omnibus level, we assess the ability of end-point reliability to approximate intersession reliability, and we compare reliability of raw and shrinkage estimates of rsFC.  We then look at seed-level reliability of raw and shrinkage estimates of rsFC to understand the reliability of seed connectivity maps for different seed regions across the brain.  For the three seeds shown in Figure \ref{fig:seeds}, we then examine the edge-level reliability of raw and shrinkage estimates to understand differences in reliability and improvement in reliability due to shrinkage across different types of connections.  Finally, we look at seed-level and edge-level maps of within-subject variance, between-subject variance, and degree of shrinkage to understand how these quantities differ across regions and connections and how they change with scan length.



\begin{figure}[t]
\centering
\includegraphics[width=6in]{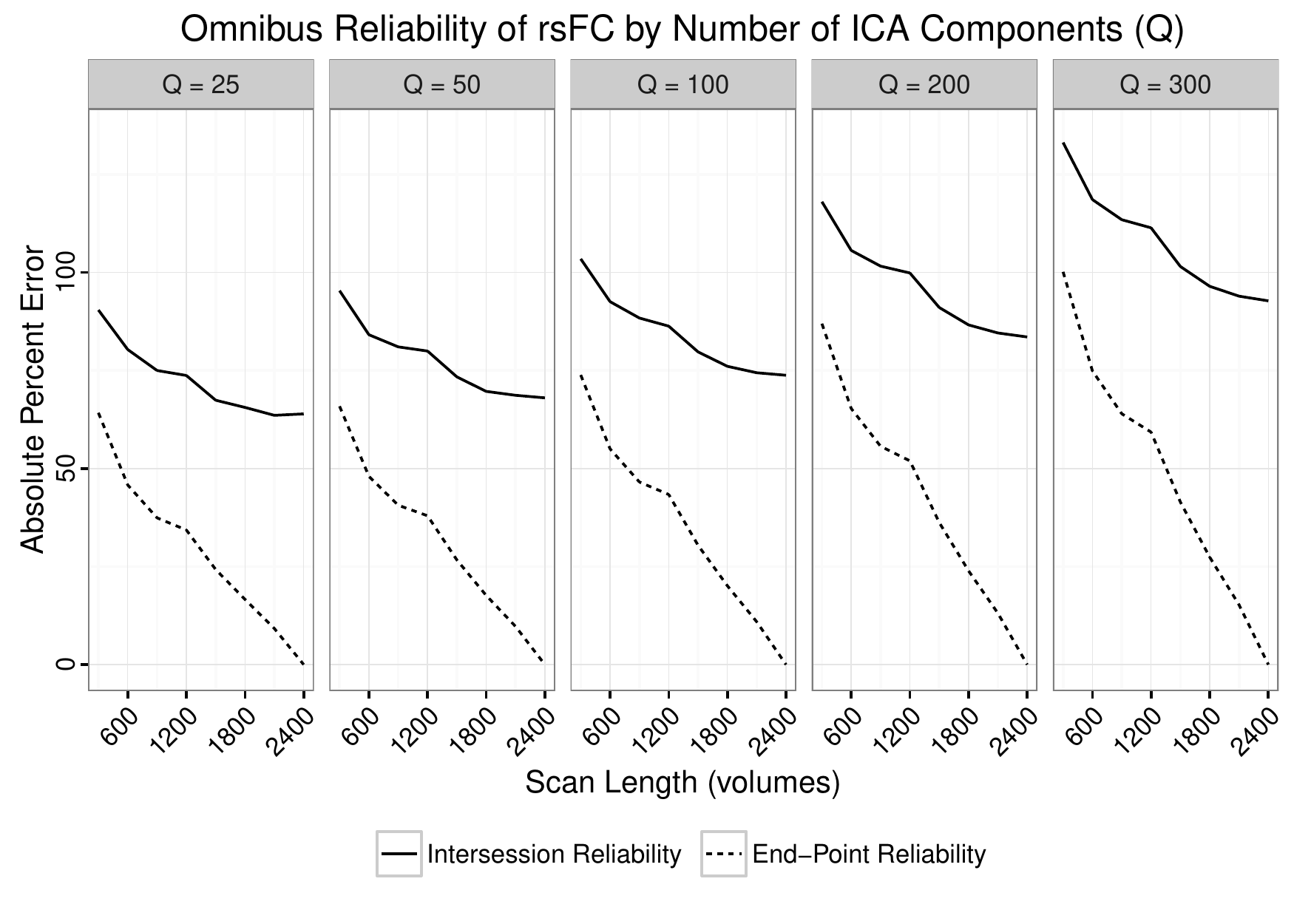}
\caption{Comparison of intersession and end-point omnibus reliability of raw connectivity estimates by scan length at each model order (25, 50, 100, 200, 300), in terms of absolute percent error.  Smaller values signify greater reliability.  For each model order and across all scan lengths, end-point error drastically underestimates the true intersession error, and this bias increases sharply as the scan length increases to $L=2400$.  This is because the estimate produced from $\ell<L$ volumes is not independent of the reference produced from all $L$ volumes, and the two quantities are equal at $\ell=L$.}
\label{fig:reliability_types}
\end{figure}

Figure \ref{fig:reliability_types} displays the intersession and end-point omnibus reliability of the raw connectivity estimates as a function of scan length and model order.  Smaller values of absolute percent error signify greater reliability.  The two measures paint very different pictures of the reliability of connectivity estimates, with end-point error drastically underestimating the true intersession error across all scan lengths.  The gap between the two measures widens as scan length $\ell$ increases and approaches $L=2400$, at which point end-point error reaches zero by definition.  This is due to the fact that, using the end-point approach to assess reliability of an estimate $W_{i1}^{(\ell)}(q,q')$, the data used to compute that estimate is a portion of that used to compute the reference $W_{i1}^{(L)}(q,q')$; thus, the two measures are not independent and, when $\ell=L$, are exactly equal.  Furthermore, Figure \ref{fig:reliability_types} shows that end-point error tends to significantly underestimate intersession error even when only a small portion of the total scan length $L$ is used to compute the estimate.  For example, at scan length $\ell=300$, end-point error underestimates the true intersession error by approximately $25$-$30\%$, depending on the model order.


Figure \ref{fig:reliability_types} also illustrates the effect of scan length and ICA model order on reliability.  Unsurprisingly, greater scan length results in improved reliability, with omnibus intersession error decreasing by approximately 30\% as scan length increases from $300$ volumes ($3.6$ minutes) to $2400$ volumes ($28.8$ minutes). However, even with nearly $30$ minutes of data per subject, reliability remains modest, with intersession APE of $60$-$100\%$, depending on the model order.  This suggests that, although increasing scan length results in more reliable estimates of rsFC, session-to-session differences in true rsFC are substantial and limit the reliability of rsFC estimates produced from a single scanning session. 

Finally, Figure \ref{fig:reliability_types} also shows that rsFC between a small number of larger regions (e.g. $Q=25$ ICs) tends to be more reliable than that of a large number of smaller regions (e.g. $Q=300$ ICs).  This may be initially surprising, as smaller regions might be expected to have more coherent signals than larger regions and hence result in better estimates of rsFC. However, there are a number of possible drivers of the observed effect.  First, the difficulty of model identification in ICA increases with the model order, and therefore there may be more error associated with estimation of a greater number of ICs.  Second, the regions were defined at a group level, and smaller regions may exhibit greater variation across subjects than larger regions.  For example, the entire somatomotor region may be spatially similar across subjects, whereas its subregions may exhibit more subject-level differences.  There may also be visit-level deviations in the spatial location of small functional regions due to minor errors in registration or normalization.  Third, the ICA time courses are essentially a weighted average across voxel-level time courses, and averaging a greater number of voxels will tend to result in reduced noise levels, and hence less noisy estimates of rsFC.  Finally, session-to-session differences in true rsFC between many smaller regions may be greater than that of a few larger regions.



\begin{figure}
\centering
\includegraphics[width=6in]{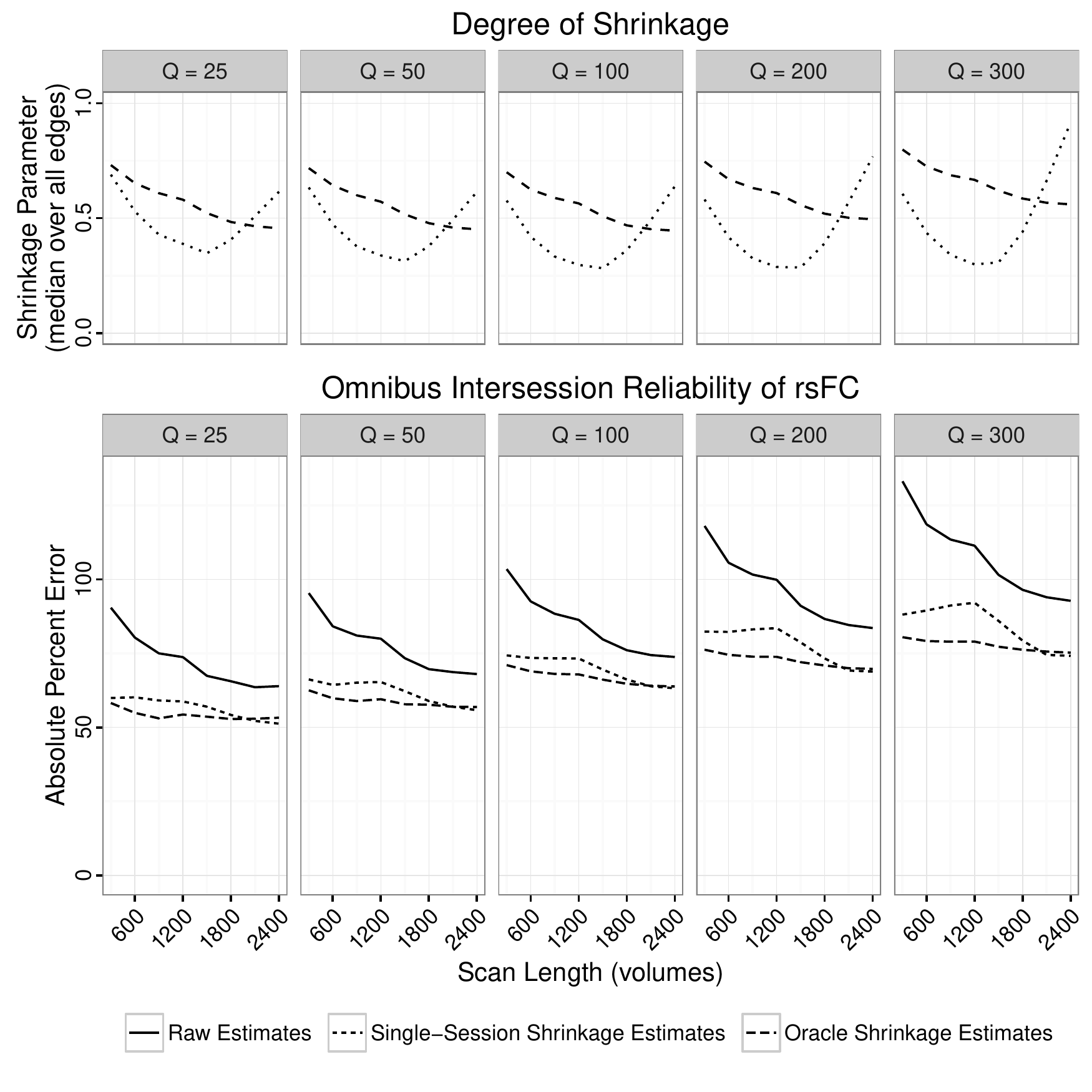}
\caption{\textbf{Top panel:} Degree of shrinkage (median over all connections) versus scan length for each ICA model order $Q=25$, 50, 100, 200, 300.  For oracle shrinkage, the degree of shrinkage tends to decrease as scan length increases.  This is expected, since raw subject-level estimates become more reliable with additional scan length and therefore require less shrinkage towards the group mean.  However, for single-session shrinkage, the degree of shrinkage exhibits an initial decrease followed by an unexpected increase.  This increase is likely due to the change in phase encoding method, which introduces an additional source of variation between the first and second half of each session.  This leads to an inflation of the within-subject variance estimated from a single session, which in turn leads to over-shrinkage.  However, this is not a flaw of the proposed single-session shrinkage methods but rather an artifact of the unique HCP acquisition protocol.  For scan lengths below $1200$ (within which there is no change in phase encoding), single-session shrinkage tends to underestimate the degree of shrinkage relative to oracle shrinkage.  This may be because true rsFC varies less within a session than across sessions.  \textbf{Bottom panel:} Comparison of omnibus intersession reliability of raw and shrinkage connectivity estimates by scan length for each ICA model order.  Smaller values of absolute percent error signify greater reliability.  Both single-session and oracle shrinkage estimates exhibit greater intersession reliability than raw estimates across all model orders and scan lengths.  Notably, shrinkage estimates produced using only $300$ volumes ($3.6$ minutes) show similar reliability to raw estimates produced using $2400$ volumes ($28.8$ minutes).}
\label{fig:reliability_shrinkage}
\end{figure}

Figure \ref{fig:reliability_shrinkage} compares the omnibus intersession reliability of raw and shrinkage estimates of rsFC (bottom panel) as well as the degree of shrinkage (top panel) as a function of scan length and ICA model order. As seen the top panel, using oracle shrinkage the degree of shrinkage decreases monotonically with increasing scan length.  This is expected, since raw subject-level estimates become more reliable as scan length increases and hence require less shrinkage towards the group mean. Using single-session shrinkage, however, the degree of shrinkage exhibits an initial decrease followed by an unexpected increase.  This increase is caused by overestimation of the within-subject variance, which is likely a result of the change in phase encoding method halfway through each session as described in Section \ref{sec:data}.  A change in phase encoding introduces an additional source of variation in estimates of rsFC.  Thus, for scan lengths $\ell>1200$, for which the last $\ell-1200$ volumes were acquired using a different phase encoding method, the within-subject variance estimated with the proposed single-session shrinkage methods is likely inflated. However, this is not a flaw of the methods per se but rather an artifact of the unique design of the HCP.  On the other hand, for scan lengths below $\ell<1200$, single-session shrinkage tends to underestimate the degree of shrinkage relative to oracle shrinkage.  This may be because true rsFC varies less within a session than across sessions.

As seen in the bottom panel of Figure \ref{fig:reliability_shrinkage}, both single-session and oracle shrinkage estimates clearly exhibit greater intersession reliability than raw estimates, and this difference is apparent across all model orders and scan lengths.  Here it is important to recall that for both raw and shrinkage estimates, the ``reference'' used to assess reliability is the \textit{raw} estimate produced from the second visit.  Notably, shrinkage estimates produced using short scans ($300$ volumes, $3.6$ minutes) display similar reliability to raw estimates produced using much longer scans ($2400$ volumes, $28.8$ minutes).  Somewhat surprisingly, oracle shrinkage estimators only marginally outperform single-session shrinkage estimators, even though, as discussed above, single-session shrinkage often over- or underestimates the appropriate degree of shrinkage.  This suggests that the benefits of shrinkage are robust to differences in the degree of shrinkage.  Notably, overshrinkage appears to have little negative impact on overall reliability.  For example, at model order $300$ single-session shrinkage results in nearly complete shrinkage to the group mean (i.e., shrinkage parameter near $1.0$) as scan length approaches $2400$, while oracle shrinkage results in approximately equal weighting of the subject-level estimate and group mean (i.e., shrinkage parameter near $0.5$).  However, the omnibus reliability of the two resulting shrinkage estimates is nearly identical.

As single-session and oracle shrinkage estimators display similar performance, in the remainder we only display the results of oracle shrinkage estimators for brevity.  While single-session shrinkage estimators are designed for use in practice, as discussed above, the unique phase encoding design of the HCP results in over-shrinkage using single-session shrinkage for scan lengths over $1200$ volumes.  In this setting, therefore, oracle shrinkage provides a realistic, albeit best-case, picture of the benefits of shrinkage.  



\begin{figure}
\centering
\includegraphics[width=6.5in]{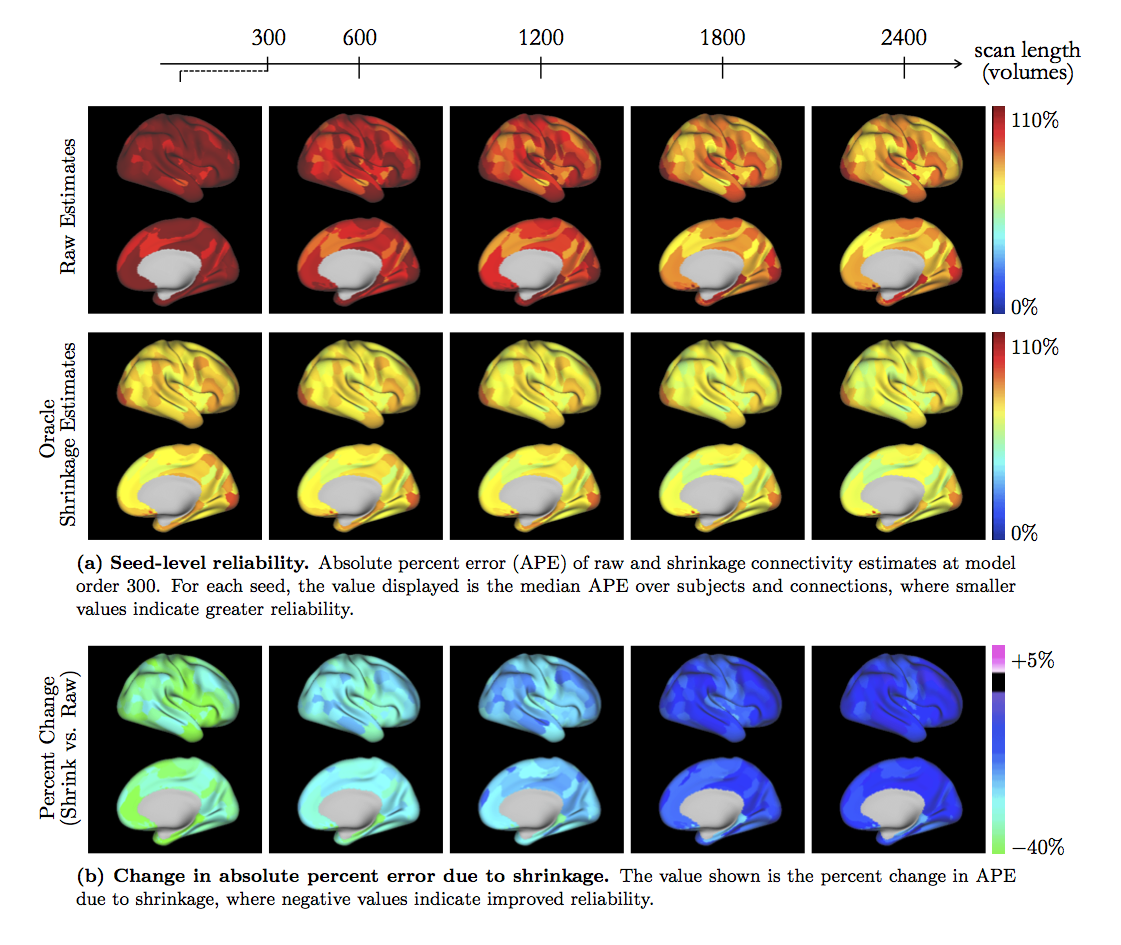}

\caption{Seed-level reliability.}
\label{fig:Reliability_seedlevel}
\end{figure}

\begin{figure}
\centering
\includegraphics[width=6.5in]{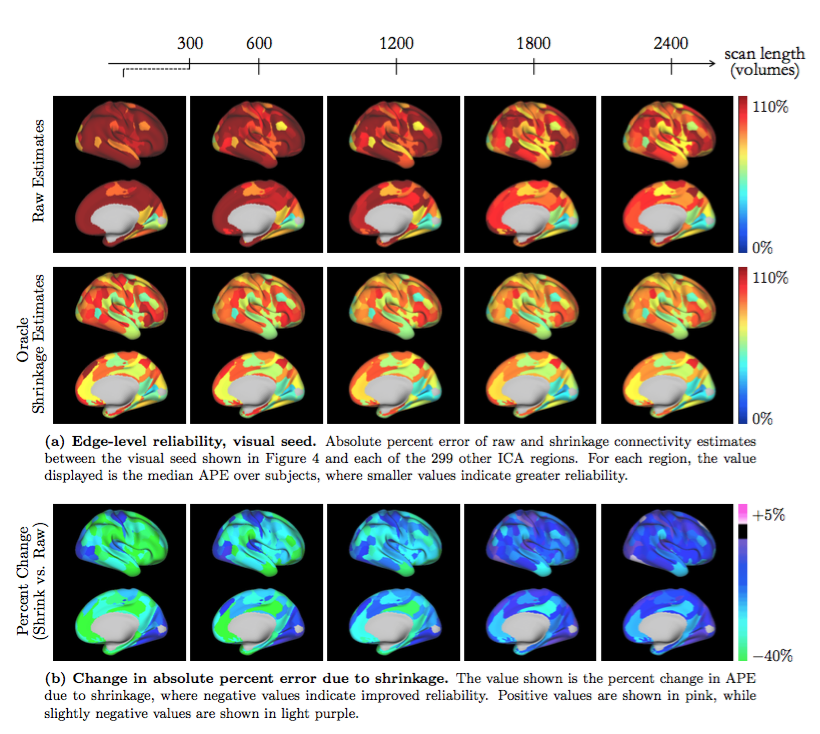}

\caption{Edge-level reliability, visual seed.  The lateral and medial surfaces of the right hemisphere are displayed.  See Figure \ref{fig:seeds} for location of visual seed (IC 16), which lies on the right medial surface and appears in gray in the images here.}

\label{fig:Reliability_visual}
\end{figure}

\begin{figure}
\centering
\includegraphics[width=6.5in]{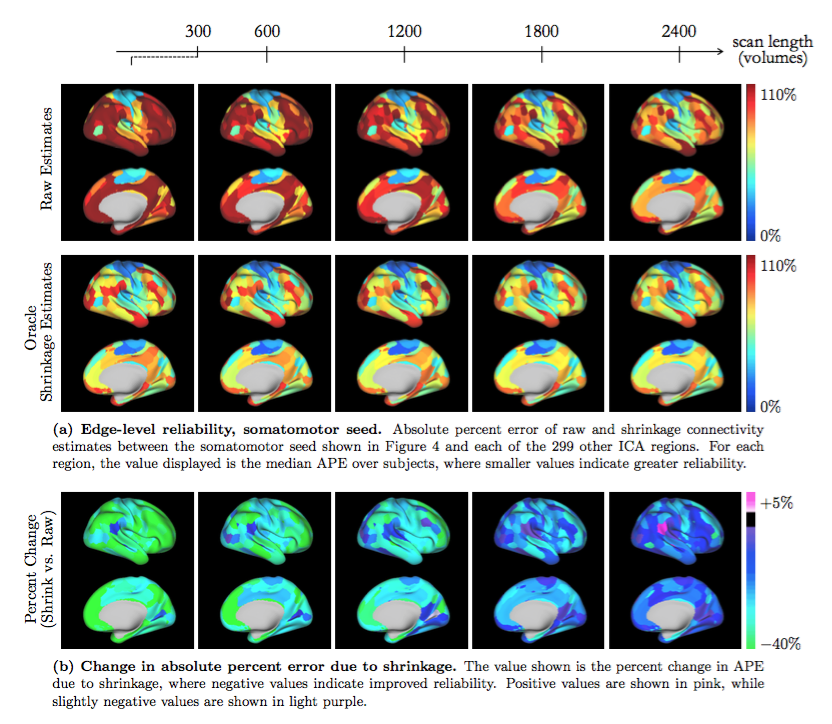}

\caption{Edge-level reliability, somatomotor seed.  The lateral and medial surfaces of the right hemisphere are displayed.  See Figure \ref{fig:seeds} for location of the somatomotor seed (IC 21), which lies on the left lateral surface and is therefore is not displayed here.}
\label{fig:Reliability_motor}
\end{figure}

\begin{figure}
\centering
\includegraphics[width=6.5in]{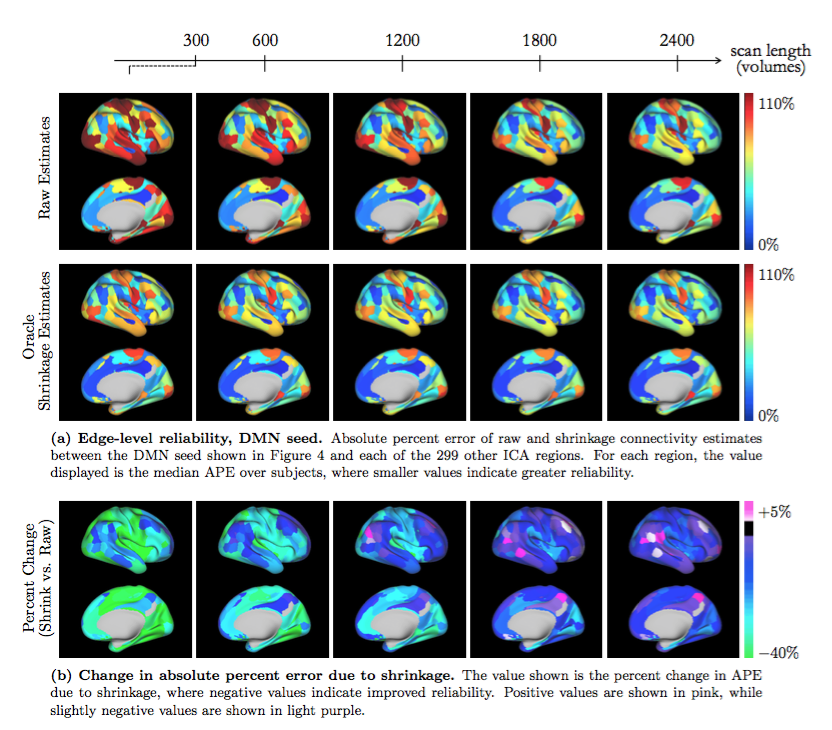}

\caption{Edge-level reliability, DMN seed.  The lateral and medial surfaces of the right hemisphere are displayed.  See Figure \ref{fig:seeds} for location of DMN seed (IC 30), which lies on the right medial surface and appears in gray in the images here.}
\label{fig:Reliability_DMN}
\end{figure}

\begin{figure}
\centering
\includegraphics[width=6.5in]{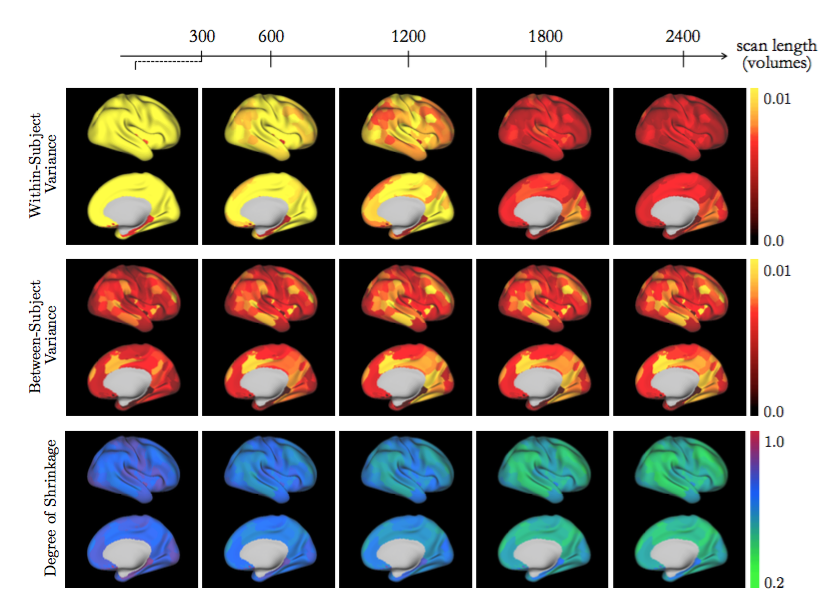}

\caption{\textbf{Seed-level variance components and degree of shrinkage.}  For each quantity, the median value over all connections with a given seed is displayed.  As scan length increases, between-subject variance stays relatively constant, while within-subject variance and hence the degree of shrinkage decrease.}
\label{fig:var_lambda}
\end{figure}

\begin{figure}
\centering
\includegraphics[width=6.5in]{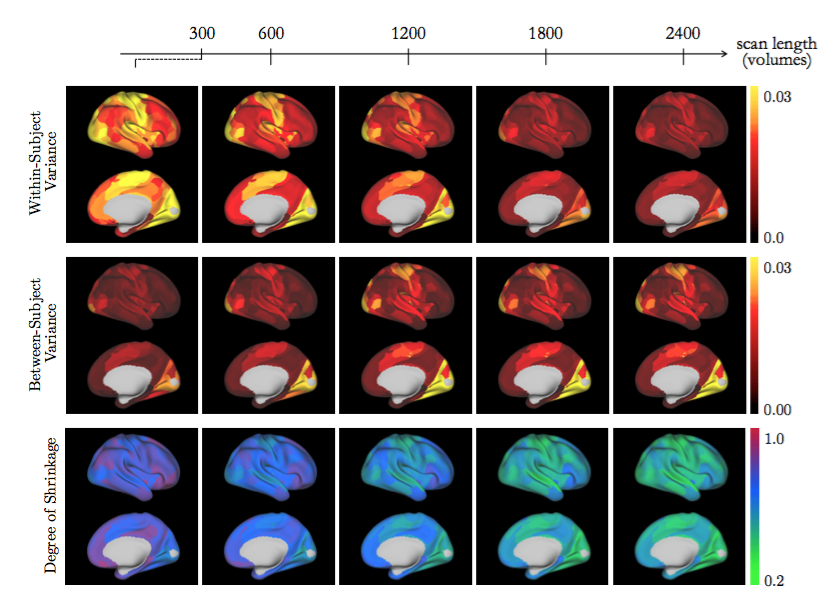}

\caption{\textbf{Edge-level variance components and degree of shrinkage, visual seed.} See Figure \ref{fig:seeds} for location of visual seed (IC 16), which lies on the right medial surface and appears in gray in the images here.}
\label{fig:var_edge_visual}
\end{figure}

\begin{figure}
\centering
\includegraphics[width=6.5in]{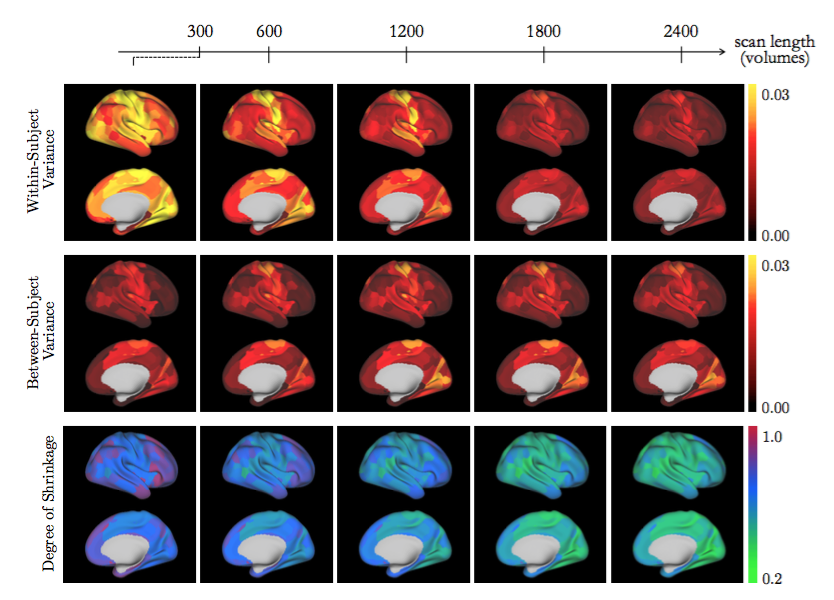}

\caption{\textbf{Edge-level variance components and degree of shrinkage, somatomotor seed.} See Figure \ref{fig:seeds} for location of the somatomotor seed (IC 21), which lies on the left lateral surface and is therefore is not displayed here.}
\label{fig:var_edge_motor}
\end{figure}

\begin{figure}
\centering
\includegraphics[width=6.5in]{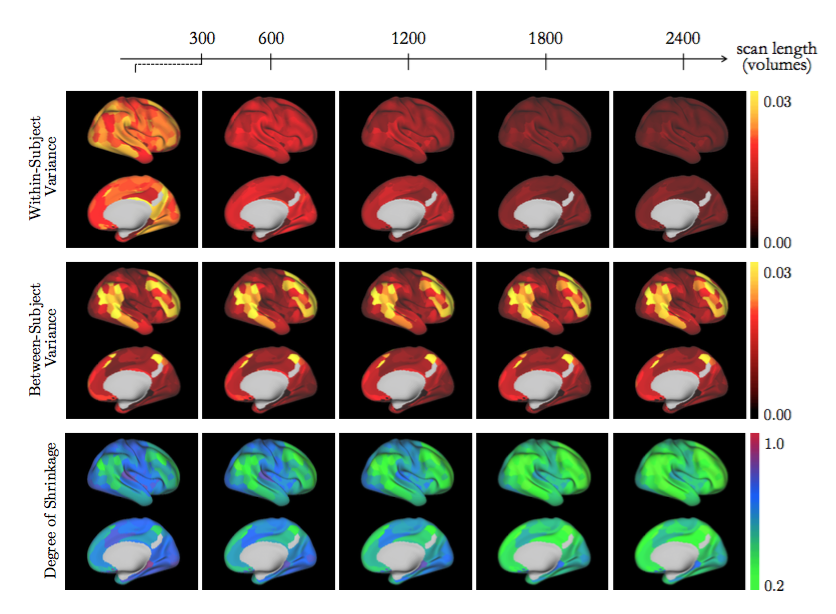}

\caption{\textbf{Edge-level variance components and degree of shrinkage, DMN seed.} See Figure \ref{fig:seeds} for location of DMN seed (IC 30), which lies on the right medial surface and appears in gray in the images here.}
\label{fig:var_edge_DMN}
\end{figure}

Figure \ref{fig:Reliability_seedlevel} displays seed-level intersession reliability maps of raw and shrinkage estimates at model order 300 as a function of scan length.\footnote{For Figures \ref{fig:Reliability_seedlevel} to \ref{fig:var_edge_DMN}, the subcortical and left-hemispheric surface grayordinates are not displayed but show similar trends.}  Recall that reliability is computed for each subject at each edge and is then summarized at the edge and seed levels as illustrated in Figure \ref{fig:reliability_hierarchy}.  Again, lower values of absolute percent error signify greater reliability.   Panel (a) illustrates that increased scan length leads to improved reliability for both raw and shrinkage estimates.  Panel (b) displays the percent change in APE after shrinkage relative to the raw estimates, where negative values indicate improved reliability, and shows that shrinkage estimates exhibit greater seed-level reliability than raw estimates for all scan lengths and seed regions.  While improvement due to shrinkage is greatest for shorter scans (approximately $30$-$40$\% decrease in APE for most regions), improvement is still substantial for the longest scans (approximately $10$-$20$\% decrease in APE for most regions).  As seen in panel (a), there is some spatial variability in seed-level reliability of rsFC.  For example, higher error is observed in some visual and motor seeds, the temporal pole and the medial temporal lobe (MTL).  However, the results shown in panel (b) indicate that the benefits of shrinkage are fairly uniform across the brain.  


Figures \ref{fig:Reliability_visual} to \ref{fig:Reliability_DMN} share the format of Figure \ref{fig:Reliability_seedlevel} but show edge-level reliability for the three seeds shown in Figure \ref{fig:seeds}.  These figures illustrate that edge-level reliability varies dramatically across connections, and within-network connections tend to exhibit greater reliability than across-network connections.  For example, Figure \ref{fig:Reliability_visual} shows that the visual seed, which is lies on the right medial surface, is most reliably connected with other medial visual regions; Figure \ref{fig:Reliability_motor} shows that the somatomotor seed, which lies on the left lateral surface, is most reliably connected with other motor seeds, including the contralateral motor regions visible in the figure; Figure \ref{fig:Reliability_DMN} shows that the DMN seed, which lies within the PCC on the right medial surface, is most reliably connected with other DMN regions, including other areas of the PCC, the medial prefrontal cortex, and the angular gyrus.  For all three seed regions, shrinkage results in improved reliability for nearly all connections and scan lengths.  Shrinkage tends to be more beneficial for less reliable connections.  For example, panel (b) of Figure \ref{fig:Reliability_visual} shows that for scan lengths up to $1200$ volumes, shrinkage results in $10$-$20$\% improvement in reliability for connections between the visual seed and other visual and motor regions and $30$-$40$\% improvement for most other connections.  However, shrinkage may also be highly beneficial for very reliable connections.  For example, Figure \ref{fig:Reliability_motor} shows that the most reliable motor connections (seen in dark blue in panel (a)) show some of the greatest improvements in reliability due to shrinkage, with $20$-$30$\% improvement even for the longest scans (see in teal in panel (b)).  In general, shrinkage results in improved reliability even for scan lengths of $1800$ or $2400$ volumes for the vast majority of connections, with up to $30$\% improvement for some connections.


Maps of within-subject variance, between-subject variance, and degree of shrinkage of connectivity estimates as a function of scan length are shown in Figures \ref{fig:var_lambda} to \ref{fig:var_edge_DMN}.  These are population-level parameters, so for each quantity there is one value per connection.  For each seed, Figure \ref{fig:var_lambda} displays the median of each quantity across all connections; Figures \ref{fig:var_edge_visual} to \ref{fig:var_edge_DMN} simply display the value of each quantity for each connection with the three seed regions shown in Figure \ref{fig:seeds}.  These figures illustrate that as scan length increases, within-subject variance tends to decrease, as subject-level estimates of rsFC become more accurate.  However, between-subject variance remains similar across different scan lengths.  Furthermore, there are clear spatial patterns of within-subject and between-subject variance, which are most apparent at the edge level.  For example, the highest between-subject variance is exhibited by connections within the DMN and visual networks (see Figures \ref{fig:var_edge_visual} and \ref{fig:var_edge_DMN}), while moderately high between-subject variance is exhibited by connections within the motor network and between the motor and visual networks (see Figures \ref{fig:var_edge_visual} and \ref{fig:var_edge_motor}).  Within-subject variance is highest for connections within and between the motor and visual networks (see Figures \ref{fig:var_edge_visual} and \ref{fig:var_edge_motor}), while within-subject variance is quite low for all connections with the DMN seed (see Figure \ref{fig:var_edge_DMN}).

High within-subject variance does not necessarily lead to a high degree of shrinkage towards the group mean since, as detailed in Section \ref{sec:conn_estimation}, the degree of shrinkage is determined by the ratio of within-subject variance to total (within-subject plus between-subject) variance.  Hence, the degree of shrinkage can be seen as a measure of reliability of raw estimates of connectivity (and is in fact equal to $1-$ICC), with lower values signifying greater reliability of subject-level estimates relative to the similarity between subjects, and therefore requiring less shrinkage towards the group mean.  Figure \ref{fig:var_lambda} shows that overall, the degree of shrinkage is lower for connections with frontal and temporal/occipital networks and higher for connections with the visual network, motor network, medial temporal lobe, and the temporal pole. Figure \ref{fig:var_edge_DMN} shows that the degree of shrinkage is lowest for connections within the DMN, due to low within-subject variance combined with high between-subject variance of these connections; Figures \ref{fig:var_edge_visual} and \ref{fig:var_edge_motor} shows that the degree of shrinkage is also relatively low for connections within and between the motor and visual networks.  Other between-network connections tend to have a higher degree of shrinkage. These observations are consistent with findings of previous studies that have examined the reliability of within- and between-network connectivity \citep{shehzad2009resting, van2010intrinsic, laumann2015functional}.

\section{Discussion}
In this paper, we examine reliability of estimates of functional connectivity between resting-state networks identified through ICA using data from the Human Connectome Project.  We investigate the effect of increasing scan length and assess the potential of empirical Bayes shrinkage to improve reliability.  We consider reliability at three different resolutions (omnibus, seed-level and edge-level) in order to study reliability of the entire connectivity matrix as well as seed maps and individual connections.  This multi-resolution approach allows us to assess the effects of scan time and shrinkage on reliability in general, as well as how these effects vary across different networks and connections.

Building upon our previous findings on the benefits of shrinkage for estimates of voxel-level connectivity within the motor cortex produced using relatively short scans \citep{mejia2015improving}, in this study we find the benefits of shrinkage to be robust to several deviations from that scenario.  Specifically, we find shrinkage to be beneficial for whole-brain connectivity between functional regions of interest produced using the high-quality, high temporal resolution data of the HCP.  We also find that while shrinkage is most beneficial for shorter scans ($30$-$40$\% improvement in overall reliability), it tends to remain beneficial as scan length increases up to (and possibly beyond) $2400$ volumes or nearly $30$ minutes ($10$-$20$\% improvement).  We find this to be the case for nearly all the connections we considered, even highly reliable connections such as those within the default mode and motor networks. 

Regarding scan length, our study suggests that, for high temporal resolution data, increases beyond $20$ minutes are likely to yield diminishing improvements in intersession reliability of rsFC.  Even with nearly $30$ minutes ($2400$ volumes) of scan time, the intersession reliability achieved by raw estimates of rsFC is quite unremarkable overall, with approximately 90\% absolute percent error at the omnibus level (for ICA model order 300) and for most specific connections considered.  Exceptions include several connections within the default mode and motor networks, which exhibit moderate to high reliability depending on scan length.  However, in general these findings suggest that session-to-session variations in rsFC are substantial for most connections, and increasing the duration within a single session only yields modest increases in intersession reliability.  This is consistent with previous findings \citep{shehzad2009resting, anderson2011reproducibility, birn2013effect, zuo2013toward} and builds on a growing body of evidence suggesting that combining data from multiple sessions, ideally occurring on different days, may result in more reliable estimates of rsFC than single-session estimates \citep{shehzad2009resting, laumann2015functional}. 

In this study, we have focused primarily on intersession reliability, which is most relevant when trait-level effects are of interest \citep{geerligs2015state}.  The proposed shrinkage methods are designed to maximize intersession reliability by taking into account both sampling variance and state-level changes in true rsFC in the estimation of total within-subject variance.  In general, state-level changes in rsFC, whether within-session or across-session, will reduce reliability of rsFC, make the connectivity measures less indicative of trait-level effects, and lead to more shrinkage.  However, if state-level rather than trait-level effects are actually of interest, the proposed shrinkage methods can be easily modified to account for sampling variance only in the estimation of within-subject variance.  In this case, the estimation of within-subject variance is greatly simplified, the degree of shrinkage will be reduced, and state-level differences in rsFC will be preserved.

Interestingly, our results show that estimates of rsFC between a small number of larger regions tends to be more reliable than that of a large number of smaller regions.  We hypothesize that this may be due to (i) increased error associated with estimation of a greater number of ICs; (ii) that smaller regions exhibit greater variation across subjects than larger regions; (iii) that averaging over a greater number of voxels results in reduced noise levels, and hence less noisy estimates of rsFC; and (iv) that session-to-session differences in true rsFC between many smaller regions may be greater than that of a few larger regions. Regardless of the ultimate reason, it is clear that the size of the regions is an important predictor for the reliability of rsFC and needs to be considered carefully.

Our results also showed a large degree of spatial variability in reliability. For example, lower reliability was found in some visual and motor seeds, the temporal pole and the medial temporal lobe (MTL). Similar variability was found when assessing both within-subject and between-subject variance.  For example, the highest between-subject variance was found in connections within the DMN and visual networks, while moderately high values were found within the motor network and between the motor and visual networks.  The within-subject variance was highest for connections within and between the motor and visual networks, while within-subject variance is quite low for all connections with the DMN. Hence, it is clear that the reliability of rsFC will depend upon the specific regions and connections of interest.

Throughout we assessed multiple types of reliability. Our results illustrate that end-point reliability is a poor proxy for true intersession reliability (see Figure \ref{fig:reliability_types}).  To realistically assess reliability in contexts where only a single session of data is available for each subject, an alternative approach would be to use split-half data.  For example, for a single fMRI session of length $L$, reliability can be assessed by estimating rsFC using the first $L/2$ volumes of the session and the last $L/2$ volumes, and comparing the two estimates. However, we do not explore this issue further in this work.

There are several limitations to our study.  First, while we propose single-session shrinkage methods for use in practice, we find that the unique design of the HCP, specifically the change in phase encoding halfway through each visit, results in overshrinkage using these methods and therefore limits our ability to assess their performance.  Second, using the HCP we are only able to assess the intersession reliability of rsFC estimates produced from scans up to $30$ minutes in duration.  While shrinkage estimates appear to improve reliability for nearly all connections for scan length up to $30$ minutes, it is difficult to predict their benefits for longer scans. 

Third, we present the median performance of raw and shrinkage estimates of rsFC, which demonstrates that shrinkage is beneficial for the majority of subjects, but does not tell us about subjects who may not benefit from shrinkage.  On a related note, the shrinkage methods we perform are connection-specific (which we recommend based on the dramatic variance in reliability across connections) but are not subject-specific.  It is quite plausible that subjects or groups differ in reliability of rsFC and would therefore benefit from differing degrees of shrinkage.  While tailoring the degree of shrinkage for different subjects is entirely plausible (and was in fact considered in \cite{mejia2015improving}), doing so increases the number of parameters that must be estimated.  This should be explored as an area of future research.  Another limitation in the proposed shrinkage methods is the assumption that variance in true rsFC across sessions can be approximated by that within a single session.  However, as intrasession differences in true rsFC tend to be smaller than intersession differences, this tends to result in undershrinkage.  While the proposed shrinkage model can be easily extended to account for this bias, the relationship between intersession and intrasession variance in true rsFC remains to be explored as a topic of future research.  

Finally, it is important to note that this study does not assess the effect of scan acquisition or preprocessing strategy on reliability of rsFC or the benefits of shrinkage.  However, previous studies have shown that the acquisition and processing methods used in the HCP tend to result in more reliable estimates of rsFC.  Therefore, we expect our findings on the benefits of shrinkage to hold or even improve in other, more common scenarios.  Furthermore, we consider only a single metric of connectivity, specifically Pearson correlation.  Other metrics, such as partial correlation or coherence, may exhibit different levels of reliability and benefits of shrinkage.



\section*{Acknowledgements}
Data were provided by the Human Connectome Project, WU-Minn Consortium (Principal Investigators: David Van Essen and Kamil Ugurbil; 1U54MH091657) funded by the 16 NIH Institutes and Centers that support the NIH Blueprint for Neuroscience Research; and by the McDonnell Center for Systems Neuroscience at Washington University.  This research was supported in part by NIH grants R01 EB016061, R01 EB012547, and  P41 EB015909 from the National Institute of Biomedical Imaging and Bioengineering, R01 MH095836 from the National Institute of Mental Health, and  the Craig H. Neilsen Foundation (Project Number 338419).

\bibliographystyle{apalike}
\bibliography{mybib}

\newpage
\appendix
\section*{Appendix: Single-Session Shrinkage Methods}

Consider a set of independent measurements $\{X_{it}\}$ from subjects $i=1,\dots,n$ at time points $t=1,\dots,T$.  Suppose that the quantity of interest for each subject is some summary statistic across time points, such as the sample mean or variance.  Let $Y_{i,\Omega}$  represent the true value of this quantity during the continuous time period $\Omega=[1,T]$, and let $\widehat{Y}_{i,S}$ represent the estimate produced using a discrete set of observations $S\subset\Omega$.  For example, let $\widehat{Y}_{i,\mathcal{T}}$ be the estimate of $Y_{i,\Omega}$ based on the full set of measurements $\mathcal{T}=\{1,\ldots,T\}$.  

We consider that the true signal can be written as $Y_{i,\Omega}=Z_i + W_{i,\Omega}$, where $Z_i\sim N\left(\mu,\sigma^2_z\right)$ is the long-term average of the subject $i$ and $W_{i,\Omega}\sim N\left(0,\sigma^2_w\right)$ is the true deviation from that value during time period $\Omega$.  We assume that $Z_i$ are independent across subjects and $W_{i,\Omega}$ are independent across non-overlapping time periods $\Omega$. For any evenly spaced sampling $S$ of $\Omega$, we consider that the estimate $\widehat{Y}_{i,S}$ can be written 
$$
\widehat{Y}_{i,S}=Y_{i,\Omega} + U_{i,S}=Z_i + W_{i,\Omega} + U_{i,S},
$$
where $U_{i,S}\stackrel{\text{ind}}{\sim}N\left(0,\sigma^2_{u,S}\right)$ and $\sigma^2_{u,S}$ depends upon the sampling $S$.  We further assume that $Z_i$, $W_{i,\Omega}$ and $U_{i,S}$ are mutually independent.

We are interested in performing empirical Bayes shrinkage on the estimate $\widehat{Y}_{i,\mathcal{T}}$, where the quantity of interest is the true long-term average $Z_i$.  There are two sources of variance around $Z_i$ associated with the estimate $\widehat{Y}_{i,\mathcal{T}}$, the \textit{signal variance} $\sigma^2_w$ and the \textit{sampling variance} $\sigma^2_{u,\mathcal{T}}$.  We must therefore estimate both within-subject variance terms as well as the \textit{population variance} $\sigma^2_z$ in order to produce the shrinkage parameter,
$$
\lambda=\frac{\sigma^2_w + \sigma^2_{u,S}}{\sigma^2_w + \sigma^2_{u,\mathcal{T}} + \sigma^2_z}.
$$
The denominator can simply be estimated as $\widehat{Var}_i\{\widehat{Y}_{i,\mathcal{T}}\}$.

\subsection*{Sampling Variance Estimation}

Without loss of generality, assume that $T$ is even and let $S_o=\{1,3,\ldots,T-1\}$ and $S_e=\{t\in 2,4,\ldots,T\}$.  Consider $\widehat{Y}_{i,S_o}$ and $\widehat{Y}_{i,S_e}$, which can be written as
\begin{equation*}
\left\{\begin{array}{lll}
\widehat{Y}_{i,S_o} & = & Z_i+W_{i,\Omega}+U_{i,S_o};\\
\widehat{Y}_{i,S_e} & = & Z_i+W_{i,\Omega}+U_{i,S_e}
\end{array}\right.
\end{equation*}
If sampling variance is inversely proportional to the number of observations in the sample (which follows by the central limit theorem for any summary statistic that can be written as a mean), then $U_{i,S_o}$ and $U_{i,S_e}$ each have variance $2\sigma^2_{u,S}$, since $S_o$ and $S_e$ each contain half the number of observations as $S$.  Observe that

\begin{align*}
Var_i\{\widehat{Y}_{i,S_o} - \widehat{Y}_{i,S_e}\}
&= Var_i\{U_{i,S_o} - U_{i,S_e}\} \\
&= Var_i\{U_{i,S_o}\}  + Var_i\{U_{i,S_e}\} \\
&= 4\sigma^2_{u,S}.
\end{align*}

Therefore, the sampling variance can be estimated as $\hat\sigma^2_{u,S}=\tfrac{1}{4}\widehat{Var}_i\{\widehat{Y}_{i,S_o} - \widehat{Y}_{i,S_e}\}$.  We note that a bootstrap approach could also be used to estimate the sampling variance with greater efficiency.  However, when enough subjects are available, the proposed subsampling approach can also result in efficient estimation.  Furthermore, the proposed approach is less computationally demanding.

\subsection*{Signal Variance Estimation}

Let $S_1=\{1,\ldots,T/2\}$ and $S_2=\{T/2+1,\ldots,T\}$, and consider $\widehat{Y}_{i,S_1}$ and $\widehat{Y}_{i,S_2}$, which can be written as
\begin{equation*}
\left\{\begin{array}{lll}
\widehat{Y}_{i,S_1} & = & Z_i+W_{i,\Omega_1}+U_{i,S_1};\\
\widehat{Y}_{i,S_2} & = & Z_i+W_{i,\Omega_2}+U_{i,S_2}
\end{array}, \right.
\end{equation*}

where $\Omega_1=[1,T/2]$ and $\Omega_2=(T/2,T]$.  Since $\Omega_1$ and $\Omega_2$ are non-overlapping, by assumption $W_{i,\Omega_1}$ and $W_{i,\Omega_2}$ are independent.  Observe that

\begin{align*}
Var_i\{\widehat{Y}_{i,S_1} - \widehat{Y}_{i,S_2}\}
&= Var_i\{(W_{i,\Omega_1}+U_{i,S_1}) - (W_{i,\Omega_2}+U_{i,S_2})\} \\
&= Var_i\{W_{i,\Omega_1}\} + Var_i\{W_{i,\Omega_2}\} +      
   Var_i\{U_{i,S_1}\} + Var_i\{U_{i,S_2}\} \\
&= 2\sigma^2_w + 4\sigma^2_{u,S}.
\end{align*}

Therefore, the signal variance can be estimated as $\hat\sigma^2_w = \frac{1}{2}\widehat{Var}_i\{\widehat{Y}_{i,S_1} - \widehat{Y}_{i,S_2}\}  - 2\hat\sigma^2_{u,S}$. 

\end{document}